\documentclass{elsart}

\usepackage{amssymb}
\usepackage{amsmath}
\usepackage{graphicx}
\begin{document}

\begin{frontmatter}

\title{NMR implementations of Gauss sums}
\author{Jonathan A. Jones}
\address{Centre for Quantum Computation, Clarendon Laboratory,
University of Oxford, Parks Road, Oxford OX1~3PU, United Kingdom}
\address{Centre for Advanced ESR,
University of Oxford, South Parks Road, Oxford OX1~3QR, United Kingdom}

\begin{abstract}
% Text of abstract
I describe the use of NMR experiments which implement Gauss sums as a method for factoring numbers and discuss whether this approach can be computationally useful.
\end{abstract}

\end{frontmatter}

% main text
\section*{Introduction}
Several papers have recently been published~\cite{Mehring, Mahesh, Peng} which describe nuclear magnetic resonance (NMR) experiments that implement Gauss sums~\cite{Merkel}, and which suggest that this can be used as a method for factoring numbers.  These experiments rely on the behaviour of the Gauss sum
\begin{equation}
\mathcal{A}_N^{M}(l) = \frac{1}{M+1} \sum_{m=0}^M \exp\left[- \mathrm{i}\,2\pi\, m^2\frac{N}{l}\right]
\label{eq:Gauss}
\end{equation}
(where $N$ is an integer to be factored, and $l$ is a trial factor), namely that $\mathcal{A}=1$ if $l$ is indeed a factor of $N$, and $|\mathcal{A}|<1$ otherwise, with the exact value depending on the choice of the truncation parameter $M$.

Clearly this behaviour allows factors to be distinguished from non-factors.  Note that this does not in itself provide a factoring algorithm, but simply a factor checking algorithm.  As with any simple algorithm of this kind, in order to factor $N$ it may be necessary to try all possible prime factors up to $\sqrt{N}$.  The number of prime numbers less than $x$ is given by the prime counting function~\cite{Ribenboim}, $\pi(x)\sim{x}/\log(x)$, but in practice it may be simpler to just consider all possible factors up to $\sqrt{N}$, or over some smaller range.  For example Mehring \textit{et al.}~\cite{Mehring} analysed the number $N=157573=13\times17\times23\times31$ by considering trial factors between 1 and 35.
%$\pi(x)=C(x)\times x/\log{x}$ with $\log(2)<C(x)<2\,\log{2}$ for $x\ge5$ (much tighter bounds on $C$ are known, but these will do for our purposes), and so the number of trial factors needed is less than $\sqrt{N}\log{16}/\log{N}$.

It is also important to consider what value of $M$ is required to reliably distinguish factors from non-factors.  In most cases it is relatively easy to distinguish these cases, but for a small number of non-factors, sometimes called \textit{ghost factors}, the Gauss sum is close to one for small values of the truncation parameter.  It can, however, be shown~\cite{Peng, Stefanak} that effective suppression of ghost factors occurs by $M=\sqrt[4]{N}$.  It is also possible to use even smaller values of $M$~\cite{Peng} either by choosing values of $m$ randomly in the range between $0$ and $\sqrt[4]{N}$, rather than simply using all the numbers in this range, or by using generalized exponential sums, a modification of Eq.~\ref{eq:Gauss} replacing the $m^2$ terms with higher powers.

\section*{NMR implementations}
Several different implementations of Gauss sums using NMR have been described, of which the conceptually simplest approach is perhaps the differential excitation method~\cite{Mahesh} of Mahesh \textit{et al.}  This uses an ensemble of identical isolated spin-$\frac{1}{2}$ nuclei in a magnetic field, which is excited by a series of small flip-angle pulses applied on resonance.  The propagator for a single pulse can be written using product operator notation~\cite{Sorensen} as
\begin{equation}
U_m=\exp\{-\mathrm{i}\theta\,[I_x\cos\phi_m+I_y\sin\phi_m]\}
\label{eq:propm}
\end{equation}
where $\theta$ is the flip angle and
\begin{equation}
\phi_m=2\pi\, m^2N/l
\label{eq:phim}
\end{equation}
is the phase of the $m^\textrm{th}$ pulse. The propagator for the whole pulse sequence is given by
\begin{equation}
U=U_M\dots U_1U_0.
\end{equation}
where in general the propagators for individual pulses do not commute and so it is necessary to use a time-ordered product. For small values of $\theta$, however, the pulses \textit{approximately commute}~\cite{Peng} so that to first order in $\theta$
\begin{equation}
U\approx\exp\{-\mathrm{i}\theta\,\sum_{m=0}^{M}[I_x\cos\phi_m+I_y\sin\phi_m]\}
\label{eq:propa}
\end{equation}
and the overall flip angle and phase of this combined pulse, and thus the magnitude and phase of the resulting NMR signal, effectively encode the value of the desired Gauss sum.  Other approaches described by Mahesh \textit{et al.}~\cite{Mahesh} and Mehring \textit{et al.}~\cite{Mehring} are essentially equivalent.

A key feature of \textit{all} current NMR implementations of Gauss sums is that they rely on the prior calculation of a set of phase angles, Eq.~\ref{eq:phim}, and all the key properties of the Gauss sum are encoded within these phase angles.  In particular a Gauss sum only achieves the value of unity when \textit{all} the phase angles are integer multiples of $2\pi$.  As $m^2$ is an integer this requires that $N/l$ is an integer.  This is, of course, precisely why Gauss sums work~\cite{Mahesh}, but it also makes clear that there is no need to explicitly evaluate a Gauss sum once a single value of $\phi_m$ is known.  Furthermore, calculating any value of $\phi_m$ requires evaluating $N/l$ to sufficient precision to reveal directly whether or not $l$ is a factor of $N$.  Thus \textit{current NMR methods to evaluate Gauss sums cannot be implemented without first implicitly determining whether or not the trial factor is indeed a factor}, in effect rendering them pointless.

It is useful to consider whether this explicit evaluation of $N/l$ can be side-stepped.  One obvious possibility is to extend Gauss sums by replacing $N/l$ by a continuous parameter $f$ and seek peaks as a function of $f$, which could then be related back to factors of $N$. It is easy to see, however, that this cannot work, as this generalisation of a Gauss sum will have peaks at \textit{every} integral value of $f$, and these only identify factors when the corresponding trial number $l=N/f$ is an integer. For example~\cite{Jones}, it is easy to distinguish $17$, which is a factor of $157573$, from 18, which is not, by using their corresponding Gauss sums.  However it is not possible to distinguish between the peak occurring at $f=9269$, corresponding to the genuine factor 17, and the equivalent peak at $f=9268$, corresponding to the non-factor $157573/9268\approx17.0018343$.

\section*{Conclusions}
Although current NMR experiments do in fact evaluate Gauss sums they cannot provide a useful method for factor testing.  Other experimental techniques have been demonstrated for evaluating Gauss sums~\cite{Gilowski, Bigourd}; it is not immediately clear whether these techniques are also subject to similar criticisms, and I do not address this point here.  However, methods based on Gauss sums will only be useful if is possible to avoid explicit division in the pre-calculation stages of the algorithm, and the integral nature of $l$ is built directly into the implementation.

\section*{Acknowledgements}
I thank the UK EPSRC and BBSRC for financial support and Ian Walmsley for helpful discussions.

\end{document}